# Picosecond laser structuration under high pressures: observation of boron nitride nanorods

*Luc Museur,[1] Jean-Pierre Petitet,[2] Jean-Pierre Michel,[2,3] Wladimir Marine,[4] Demetrios Anglos,[5] Costas Fotakis,[5,6] Andrei V. Kanaev [2#]*

[1] Laboratoire de Physique des Lasers - LPL CNRS, Institut Galilée, Université Paris 13, 93430 Villetaneuse, France

[2] Laboratoire d'Ingénierie des Matériaux et des Hautes Pressions - LIMHP CNRS, Institut Galilée, Université Paris 13, 93430 Villetaneuse, France

[3] Laboratoire des Propriétés Mécaniques et Thermodynamiques des Matériaux - LPMTM CNRS, Institut Galilée, Université Paris 13, 93430 Villetaneuse, France

[4] Centre de Recherche en Matière Condensée et Nanosciences de Marseille - CINaM, UPR CNRS 3118, Université de Marseille, 13288 Marseille, France

[5] Institute of Electronic Structure & Laser, Foundation for Research and Technology - Hellas (IESL-FORTH), 71110 Heraklion, Crete, Greece

[6] Department of Physics, University of Crete, Heraklion, Crete, Greece

---

[#] Correspondent author. E-mail: kanaev@limhp.univ-paris13.fr

**Abstract**

We report on picosecond UV-laser processing of hexagonal boron nitride (hBN) at moderately high pressures above 500 bar. The main effect is specific to the ambient gas and laser pulse duration in the ablation regime: when samples are irradiated by 5 ps or 0.45 ps laser pulses in nitrogen gas environment, multiple nucleation of a new crystalline product - BN nanorods - takes place. This process is triggered on structural defects, which number density strongly decreases upon recrystallization. Non-linear photon absorption by adsorbed nitrogen molecules is suggested to mediate the nucleation-growth. High pressure is responsible for the confinement and strong backscattering of ablation products. A strong surface structuring also appears at longer 150-ps laser irradiation in similar experimental conditions. However, the transformed product in this case is amorphous strongly contaminated by boron suboxides $B_xO_y$.

# 1. Introduction

A research on ultra-short laser modification of materials has been a subject of numerous studies last two decades. An important issue is related to excited-state mediated transitions specific to high laser fluences. Related studies are practically limited to amorphization of a crystalline phase often called "cold melting" [1-3]. Its dynamics can be directly observed by time-resolved X-ray diffraction, though more often less sophisticated pump-probe reflectivity methods are employed for the analysis of fast structural transformations. The underlying non-classical mechanism of the melting consists in solid matrix destabilisation after a transfer of ~15% bound valence-band electrons of a semiconductor to the conduction band. This mechanism is relevant to the range of laser pulse durations shorter than characteristic time of electron-phonon relaxation: $\tau_L \leq \tau_{e-ph} \approx 1\text{-}10$ ps. Under these conditions, the order-to-disorder transition can be observed. After the electronic energy is transferred to the lattice, the amorphous state is attained. Recrystallization of a sample after irradiation depends on the heat transfer and requires much longer time: $\tau_{cr} \gg \tau_{e-ph}$. It is *a priori* not expected to depend on $\tau_L$ for ultrashort laser pulse durations. However, the electronic excitation-enhanced crystallization of solids has been observed in few studies [4-7]. Solis et al. [4] have firstly reported a crystallisation threshold decrease of GeSb target material at laser pulse duration shorter then 0.8 ps; the contribution from non-linear absorption has been discarded as responsible for this effect. A short-lived highly non-equilibrium state and defects formed during the electron-phonon exchange are suggested to be a driving force of subsequent rapid nucleation of a new crystalline phase [8].

Practically, attending the critical electronic excitation for femtosecond pulse durations often requires laser fluences of ~$10^2$ mJ/cm² or higher that is comparable to a material ablation threshold [9]. Under vacuum, the ablated products leave the target escaping analysis. Many interesting information related to influence of electronic excitation of the ablated products on nucleation process can be lost in these conditions. On the other hand, ablation products can be confined by applying high pressure of the surrounding gas. A backscattering of the ablation products can contribute to a specific surface structuring [10]. Typically, the relevant pressure is about 100 mbar for argon or nitrogen gases whereas higher for light gas helium (~1 bar). However, in these conditions only the ground-state products subject to scattering events. Higher pressures may stronger confine the ablation plume forcing the excited-state products to scatter at the Mach disc. Taking the characteristic excited-state lifetime of ~1 µs and the molecular/atomic velocity of ~$10^3$ m/s, the pressures of ~$10^3$ bar would be required. The scattering of the excited-state products may change a general picture of the nucleation-growth process as well as allow a laser control of the surface morphology.

In this work we report on ps-laser assisted nucleation of BN nanorods from hexagonal boron nitride (hBN) powder. The hBN samples are irradiated by picosecond UV lasers above the threshold of thermal ablation ($E_a \leq E_L \leq 10 E_a$, $E_a$=80 mJ/cm$^2$) at moderately high nitrogen gas pressures $p \leq 10^3$ bar and in presence of traces of oxygen (~1 mbar). We demonstrate a net difference in both nature and morphology of the samples obtained by employing laser pulse durations of 150 ps on one hand and 5 ps / 0.45 ps on the other.

## 2. Experiment and sample preparation

The target samples were compacted from hBN powder (Aldrich – 325 mesh) under the pressure of 0.6 GPa in square pallets (8x8x1 mm). To avoid organic impurities and traces of water, the pallets were heated at 800 K during 12 hours. The grit size of the hBN powder used for the experiment has been estimated by granulometry and transmission electron microscopy (JEM 100C JEOL). It ranged from 0.3 to 10 μm with an average particle size of 3.1 μm corresponding to the maximum in the mass distribution curve. Using conventional X-ray powder diffraction, the lattice parameters of hBN are determined $a$ = 0.2504 nm and $c$ = 0.6660 nm.

The laser experiments were carried out at Ultraviolet Laser Facility operating at FORTH (Heraklion, Greece). The hBN samples were irradiated by picosecond lasers at fluences between 1 mJ/cm² and 1 J/cm², under primary vacuum (~$10^{-2}$ mbar) and in different environments of $N_2$, $O_2$, and Ar gases (p<$10^3$ bar) and at room temperature. The laser pulse duration were 150 ps ($\lambda_L$=266 nm), 5 ps (248 nm) and 0.45 ps (248 nm). The samples were characterized by time-resolved fluorescence, Raman, XRD, TEM, and EELS methods.

The hBN samples were placed into a specially designed optical cell described in Ref. [11], which operates in the pressure range between $10^{-3}$ mbar and $10^3$ bar. Input and output cell windows are made of sapphire that is a transparent material in the UV spectral range λ≥200 nm.

Fluorescence and fluorescence excitation spectra in the UV-VUV spectral region of fresh and irradiated hBN samples were measured at the Superlumi experimental station of Hasylab of synchrotron DESY [12-13]. During these measurements the samples were maintained at low temperature of 9 K under high vacuum of ~$10^{-9}$ mbar. The samples were excited by tuneable synchrotron radiation (Δλ = 0.32 nm) in the UV-VUV spectral region.

Raman spectra have been measured using a micro-Raman high-resolution HR800 installation from Jobin Ivon Horiba with the spectral and spatial resolution correspondingly 0.25 cm$^{-1}$ and ~5 μm. The scattered light is collected by Peltier-cooled CCD camera in backscattering configuration. Most of the measurements were conducted at 488 nm using argon ion laser. However, selected series were also done at 633 nm, what is specified in the text.

The TEM is a 200kV electron gun from JEOL (2011 high resolution) with an emission type $LaB_6$ (field emission). The resolution point to point is 2.3 Å. Complimentary EELS measurements have been performed with a EELS/GIF device (GATAN IMAGINE 2000). The resolution of the energy filter is 1 eV and the dimensional resolution is 1 nm.

## 3. Results

The electronic and crystalline structures of hBN have been extensively studied last decades because of its multiple large-scale applications [14]. In contrast to earlier works, its band-gap energy has been revised to a higher value of ~6 eV in recent experimental [12-13,15] and theoretical [16] studies. The impurity states and structural defects of hBN are now characterized: in particularly, they are responsible for a strong absorption bands above 4.0 eV. Use of UV lasers is therefore preferable for hBN transformation, because 2-photon transitions involving lowest-order nonlinearity as well as 1-photon transitions to impurity states may efficiently assist the excitation process.

### *3.1 Laser irradiation*

Irradiation produces different effects on hBN samples depending on laser wavelength, pulse duration, and fluence. In the nanosecond domain of pulse durations the laser photons of energy higher than 2.57 eV promote the B-N bonds breaking:

$$BN(s) \rightarrow B(s) + 1/2 N_2(g) \tag{1}$$

forcing nitrogen to leave the sample surface [17]. However as we have recently shown, in presence of oxygen the produced defect sites are efficiently healed [18]. In mild irradiation conditions, only the uppermost surface layer is subjected to chemical modifications. Because of this, the healing prohibits the sample erosion and boron enrichment terminates. Our results indicate that oxygen reacts with multiple nitrogen vacancies ($V_N^k$, k≥2) at hBN surface and reversibly physisorbed at single nitrogen vacancies ($V_N$). The healing involving $V_N^2$ then proceeds in following steps:

$$V_N(s) + O_2 \leftrightarrow -B...O_2 \qquad (2)$$

$$V_N(s) \xrightarrow{h\nu} V_N^2(s) \qquad (3)$$

$$V_N^2(s) + \frac{1}{2}O_2 \rightarrow ...-(B-O) \qquad (4)$$

The reactions involving larger clusters vacancies are less probable because of the $V_N^2$ healing (4). The surface healing results in creation of suboxide species $B_xO_y$. This is an example of inherently photochemical process taking place in its pure state at low laser fluences.

We have observed surface darkening under picosecond UV irradiation of hBN in primary vacuum conditions. In contrast, in the ambient atmosphere no surface darkening appears at doses as high as $10^3$ J/cm². In the same time, the characteristic hBN fluorescence of irradiated samples is not attenuated [18], which indicates extremely thin oxidized layer. In the ablation regime ($E_a \leq E_L \leq 3E_a$) only a surface modification due to a mass loss has been observed.

In contrast, the experiments performed under high gas pressure p≥500 bar (nitrogen, argon, or neon) evidence strong surface structuring that is shown in Fig. 1(a-c). The observed structure appears as a result of surface instability and is related to the mass transfer and/or changes in chemical composition. The measured threshold of the structure appearance ($E_{str}$) is found almost equal to the ablation threshold ($E_a$)

$$E_{str} = E_a \qquad (5)$$

which suggests that the ablation mechanism is responsible for the structuring. The validity of Eq. (5) has been verified for irradiation of hBN samples at different wavelengths of 266 nm, 248 nm, and 213 nm. Moreover, Eq. (5) is held for 150 ps, 5 ps, and 0.45 ps pulse durations. Earlier, the threshold of hBN thermal ablation has been measured in the femtosecond domain of laser pulse durations (248 nm, 0.45 ps, single-shot regime): $E_a$=80 mJ/cm² [17]. The fact that $E_{str}$ does not appreciably changes in the considered range of the laser pulse durations indicates that the energy dissipation due to thermal conductivity of samples is low at t≤150 ps.

In presence of 1-20 mb oxygen and pressure above 500 bar, the samples are totally oxidized exhibiting white colour. On the other hand if oxygen partial pressure in the ambient gas is below 0.2 mbar, the dark coloration appears. The irradiated samples in these conditions are composed of a mixture of black and white zones (Fig. 1). Moreover, the observed picture is dramatically changed if the sample in contact with $N_2$ gas under pressure is irradiated with shorter 5-ps laser pulses (b) instead of 150-ps ones (a). A very intense surface structuring appears, which grows against the incident laser beam. The height (h) of this structure increases with gas pressure: we observed h~80 μm at 500 bar and h~480 μm at 800 bar. In contrast, the morphology of the structure is different from that obtained at both longer 150 ps (a) and shorter 0.45 ps (c) laser pulse durations. While in the former case the irradiated area contained dispersed microparticles, in the latter one the structure is more homogeneous but of lower height. The strongest surface structuring has been observed at laser fluences $2E_{str} \leq E_L \leq 3E_{str}$. At higher fluences the structuring weakens.

*3.2 Raman analysis*

In agreement with our earlier studies of nanosecond laser irradiation [17-18], the surface darkening is assigned to elementary boron. Amorphous boron *a*-B is known to exhibit broadband

Raman spectra situated in the low-frequency range below 1200 cm$^{-1}$ [19]. However as shown in Fig. 2, a very broad band at 1820 cm$^{-1}$ dominates the spectra recorded in our hBN samples irradiated in presence of oxygen (p≥0.2 mb). In particularly, the micro-Raman spectrum from the black zone of the sample from Fig. 1a is similar to that shown in Fig. 2a. Moreover, this band even more intensifies as the oxygen content increases in the ambient atmosphere while the characteristic hBN peak at 1367 cm$^{-1}$ ($E_{2g}$) vanishes. Relevant stable products, hBN ($E_{2g}$: $\omega$=1367 cm$^{-1}$) [20-21] and $B_2O_3$ ($\omega$=808 cm$^{-1}$) [22], exhibit intense sharp spectral features. Additionally, our spectra of are different from those of amorphous $B_2O_3$ [22] and $B_6O$ [23], which are the only studied compound of the boron oxide family. We therefore suppose that the observed broad bands in this spectral range are inherent to amorphous suboxide phase $B_xO_y$. Similar effect of laser picosecond processing on microcrystalline graphite is known [24]. This conclusion about amorphous nature of the modified material is supported by XRD pattern (not shown).

The $B_xO_y$ is observed at long 150-ps pulse hBN irradiation (Fig. 2a) at all used laser fluences. Apparently, a presence even of oxygen traces is sufficient for a formation of a thin boron suboxide layer, which is maintained in the ablation regime prohibiting bulk oxidation. At the irradiation below ablation threshold the suboxide species are abundant also for shorter laser pulses (Fig. 2b). However, under combined conditions of

- high nitrogen pressure p≥500 bar,
- low oxygen pressure p<0.2 mbar,
- ablation regime,
- short laser pulses $\tau_L$≤5 ps

new species appear. This can be evidenced from Raman spectra of the irradiated zone that change when laser fluence exceeds the ablation threshold (see Fig. 2c for 0.45 ps laser pulse duration).

The boron suboxide band weakens accompanied by an appearance of a new narrow band at $\omega_x$=1590 cm$^{-1}$ and with spectral width (full width half maximum, FWHM) $\Delta\omega_{1/2}$=75 cm$^{-1}$. In the same time, the hBN peak is slightly broadened and red-shifted. The shift becomes stronger when the $\omega_x$-band intensity increases and may attain $\Delta$=-15 cm$^{-1}$. In the same time, the peak width at the half maximum intensity increases from 9 cm$^{-1}$ to 18 cm$^{-1}$. This behaviour may reflect the appearance of nanometric-size species. Moreover, our results show that the nitrogen gas is essential for their appearance: when nitrogen was replaced by neon, the spectra looked similar to those of the oxidized hBN samples (Fig. 2d).

A validity of our observations in irradiated hBN samples has been checked by complementary Raman measurements at 633 nm. Surprisingly, they did not show the broad band at 1820 cm$^{-1}$ assigned to $B_xO_y$: the correspondent spectra were flat containing only the $E_{2g}$ peak at 1367 cm$^{-1}$ (hBN). This finding allows supposing a fluorescence origin of this band. It may be also Raman band requiring particular excitation photon energies. A definite conclusion about its nature may be given in future studies. In contrast, the appearance of the new $\omega_x$=1590 cm$^{-1}$ band in the specific irradiation conditions was confirmed in these Raman series.

Let us now zoom to the spectral range between 1250 cm$^{-1}$ and 1750 cm$^{-1}$, where Raman bands under interest are positioned. The band at ~1586 cm$^{-1}$ ($\Delta\omega_{1/2}$=48 cm$^{-1}$) attributed to $E_{2g}$ stretch mode of carbon has been earlier reported in $B_xC_yN_z$ nanotubes [25]. However, this cannot explain our spectra. First as Fig. 3 shows, the measured lineshape is considerably different from that of a 2D graphite sheet and the spectral width larger. Second, the carbon impurity is very low in our hBN samples and cannot provide $B_xC_yN_z$ nanotubes. Moreover, the same Raman band has been observed under irradiation of pyrolytic hBN sample in identical experimental conditions, where because of the synthesis method the carbon impurities are evanescent.

Different authors have earlier reported on a small blue shift of the main $E_{2g}$ Raman band in BN nanotubes [25-26]. This shift is related to a weak interaction between the hexagonal sheets of bulk hBN resulting in an elongation of the B-N in-plane bonds and consequently, softening of the phonons. In framework of this model, the strong red shift of the $E_{2g}$ band can be explained by even stronger phonon softening. This softening is due to a local sample heating by Raman laser at 488 nm, which is explicitly shown in the lower part of Fig. 3. The peak shifts here by up to 15 cm$^{-1}$ as cw-laser power increases from 4 mW to 40 mW. We remark that the Raman spectra of BN nanotubes (shown in Fig. 3) show no spectral shift when laser power in identical conditions varied between 4 mW and 50 mW. In agreement with earlier publications [25-26], the $E_{2g}$ peak in this case is positioned at slightly higher frequency by $\Delta=+1.5$ cm$^{-1}$ compared to bulk hBN sample. The red shift has not been observed in the hBN reference sample even at the maximum laser power of 50 mW. Also, the hBN samples irradiated with pulse duration of 0.45 ps in presence of oxygen (Fig.2b) or in absence of nitrogen (Fig.2c) show no red shift of the $E_{2g}$ Raman band.

In addition, we observed a correlation: as soon as the $\omega_x=1590$ cm$^{-1}$ band appears the $E_{2g}$ phonon mode displays a high sensitivity to the cw-laser power. We believe that this effect originates from the sample nanostructuration. A very small heat capacity of the produced nanometric species explains this increased sensitivity to heating. R. Areal et al. [26] have suggested this effect to occur in multiwall BN nanotubes. However, relevant species nucleated in our experiments are different from multiwall nanotubes, as displaying much higher thermal sensitivity (see in Fig. 3). In following we identify them with hBN nanorods.

Additional confirmation of this hypothesis has been found by monitoring Raman spectra across different treated samples. In such a way, objects of different size could be examined. We have observed that the stronger band sensitivity correlates to its spectral width. For example, the shift $\Delta\omega=4$ cm$^{-1}$ corresponds to FWHM=10 cm$^{-1}$ while $\Delta\omega=15$ cm$^{-1}$ corresponds to FWHM=14 cm$^{-1}$. This width (in cm$^{-1}$) can be related to the nanocrystallite size L (in nm) as suggested by Nemanich et al. [27]: $FWHM = 141.7/L + 8.7$. By using this formula, we estimate the size of the abundant species as correspondingly 100 nm and 27 nm. Crystallites of a larger size possess higher heat capacity and are less sensible to cw-laser intensity as shows the smaller shift of the $E_{2g}$ Raman band.

We have observed the similar effect on Raman spectra of hBN samples with 5-ps KrF-laser irradiation above the ablation threshold in identical environmental gas conditions: the band $\omega_x=1590$ cm$^{-1}$ appears accompanied by an increased sensitivity of the $E_{2g}$ band position to the cw-Ar$^+$ laser power. The correlation between the red shift of the $E_{2g}$ band and the $\omega_x$-band intensity was thus confirmed. In addition, both larger bandwidth and stronger sensitivity to Raman cw-laser power were observed in hBN samples irradiated by 0.45 ps pulses as compared to 5-ps irradiated samples. This suggests that irradiation with longer KrF-laser pulses results in generally larger size of nanorods.

### *3.3 TEM analysis*

TEM images of the hBN target before and after UV-laser irradiation with 5-ps pulses are shown in Fig. 4. Although we discuss in the following images of the samples exposed to 5-ps laser irradiation, similarly results were also obtained with 0.45-ps laser irradiation.

The raw powder-like material consists of mono-crystalline grains (a) showing characteristic electron diffraction pattern of hexagonal boron nitride (b). After irradiation, some grains melt (c) and the electron diffraction pattern weakens in intensity and shows a poly-crystalline sample nature (d). This can be understood if larger magnification image is taken from the grain (c): one can see a multiple nucleation of the new product - nanorods (e). Their diameter is between 15 and 30 nm (we remark that this corresponds to the crystallite size L estimated above from the Raman spectra); depending on irradiation conditions they grow attaining a length of several microns (f). The crystalline nature of an individual nanorod (from left down part of Fig. 4f) is evidenced in Fig. 5. The crystalline planes of this unit are distinguished in the high-resolution image (a). The rods are monocrystalline and the accompanying electron diffraction pattern (b) shows structural periodicity with $d_1$=3.34 Å corresponding to the hkl=002 plane of the raw hBN material.

The EELS analysis of the irradiated zone (correspondent to the left lower part of Fig. 4f) is presented in Fig. 6. The TEM image (left) shows a small nanorod of diameter 15 nm and length 200 nm on the hBN surface, surrounded by a cloudy material. The elemental maps indicate that the cloudy material belongs to elementary boron. On the other hand, the nanorod shows a similar composition to that of the raw material. No carbon contamination of samples subjected to laser irradiation and only weak surface contamination by oxygen are observed. This allows concluding about boron nitride origin of the observed nanorods.

*3.4 Luminescence analysis*

The characteristic low-temperature fluorescence (F) and fluorescence excitation (E) spectra of hBN are shown in Fig. 7a. According to theoretical study [16] and in agreement with experimental observations [12-13,15], the strongest band in excitation spectra at 5.96 eV (208.0 nm) is due to merged excitonic transitions with unusually large oscillator strength, while the direct band gap energy is higher. On the other hand, the lower-energy bands above 4.09 eV belong to impurity and defect absorption. The structured fluorescence band between 300 nm and 350 nm (F) may be assigned to carbon impurities and nitrogen vacancies and its complex lineshape is explained by phonon replica [18]. Earlier, it has been suggested of using the 4.0-eV fluorescence as a measure of the sample purity. The low-temperature fluorescence and fluorescence excitation spectra of hBN samples after irradiation by 5-ps laser UV pulses at 500 bar of $N_2$ and $E_L>E_a$ are shown in Fig. 7b. The fluorescence excitation bands at 5.96 eV and higher energies related to the intrinsic electronic transitions in hBN remain strong after irradiation. On the other hand, those at lower energies (5.97 eV< E ≤ 4.09 eV) related to impurities as well as the impurity fluorescence F vanish. These finding strongly supports our conclusion about advanced purity of the crystallized BN nanorods.

## 4. Discussion

Below we will briefly discuss the nature of the observed laser-induced surface structuring.

Our experiments show a definite threshold character of the hBN transformation by UV laser irradiation. The threshold for a pulse duration of 150 ps in the multiple-shot regime $E_{150ps}=10^2$ mJ/cm² corresponds to the single-shot ablation threshold. We observed that in these conditions only the uppermost layer of width of several micrometers is affected by amorphization. This amorphous phase consists of chemically transformed suboxide $B_xO_y$, while the deeper layers remain hBN. A simple estimation shows that the maximal depth on which the hBN can be affected is $\delta \approx E_L/\rho C_p \Delta T$ ~1 μm (here $\rho_{hBN}$=2.34 g/cm³ is the hBN density, $\Delta T = T^*_{hBN} - T_{room}$ where $T^*_{hBN} \approx$ 3000 K is the decomposition temperature of hBN, and $C_p \approx$ 0.8 J/(g K) is the specific heat capacity at constant pressure of hBN at room temperature [28]). This estimation seems reasonable as the laser energy flux during the longest irradiation time of $\tau_L$=150 ps ($q_L = E_L/\tau_L \approx 10^9$ J/cm²s) is much higher compared to the thermal dissipation flux ($q_T = \chi \cdot grad(T) \approx \chi \Delta T/\delta \approx 3\cdot 10^7$ J/cm²s, where $\chi \leq$ 6 W/(cm K) stands for the hBN thermal conductivity coefficient [29]): $q_L >> q_T$. We remark that boron oxides $B_xO_y$ may melt at considerably lower temperatures than hBN (for example $T_m$=450 °C for $B_2O_3$). The fact that deeper layers are affected can be explained by a melt phase penetration into the porous target.

The most interesting feature of the present study concerns crystallization of pure hBN phase of nanorod morphology. Both threshold fluences for amorphous $B_xO_y$ and crystalline BN nanorods appearance are similar within the experimental error bars to the ablation threshold. However, 150-ps pulse duration results in the amorphous phase while the crystallites reappear after 5-ps and shorter pulse irradiation. We relate this effect to participation of laser-excited products.

First we consider the effect of the pressure. During the laser irradiation of samples above the ablation threshold ionized, excited and neutral states leave the surface forming ablation plume. A confinement of the expanding plume takes place at the Mach disc, which distance from the target sensitively depends on the ambient gas pressure: $l_{Mach} = 2/3 \cdot d \sqrt{p_0/p_b}$ [30], where d is the spot diameter, and $p_0$ and $p_b$ are the initial and background gas pressure. At this point the adiabatic expansion terminates and the ablation products subject to scattering in collisions with ambient gas. From a general point of view, the excited products may appreciably contribute to the solid nucleation process if their time-of-flight ($t_{TOF}$) from the scattering zone to the sample is shorter then the excited-state lifetime:

$$t_{TOF} = l_{Mach}/\overline{V} \leq \tau^* \qquad (6)$$

where $\overline{V}$ is mean expansion velocity. The pressure ratio can be expressed as $p_0/p_b = n_{BN}T_0/n_b T_b$ ($n_{BN}$ and $n_b$ are the molecular concentrations of solid and gas) and the mean expansion velocity $\overline{V} = \sqrt{2kT_0/m_{BN}}$ ($T_0=T^*_{hBN} \approx$3000 K, $T_b$=300 K, $m_{BN}$=24.8). Using $n_{BN} = \rho/m_{BN}$, the relation (6) can then be reduced to:

$$d\sqrt{\frac{2\rho}{9kn_bT_b}} \leq \tau^* \qquad (7)$$

where ρ=2.18 g/cm³ is the BN specific mass and k=1.38·10⁻¹⁶ erg/°C. In the experimental conditions of d=1 mm and ambient pressure of $p_b$ =10³ bar ($n_b$ =2.5·10²² cm⁻³) the excited-state lifetimes under concern is $\tau_{TOF}$ ≥1 μs. The scattering zone distance $l_{Mach}$ ≤800 μm can define the characteristic height of the surface structuring, which is in a general agreement with the experimental observations.

The most probable metastable species participating in the recrystallization process are N atoms generated by photolysis of N₂ gas by 248 nm laser photons:

$$N_2 \xrightarrow{h\nu} 2N \qquad (8)$$

However, no electronic states of N₂ available to dipole-allowed one- and two-photon transition exist in this spectral range [31]. Although higher nonlinearity may also contribute to absorption in this spectral range (as for example 4-photon transition resulting in population of the $N_2^+(B^2\Sigma_u^+)$ state [32]), their probability is expected to be rather low. Our experiments show that the transmission of the high-pressure nitrogen at $E_L = 2E_a$ and 500 bar of N₂ is about 70% that can be explained by this non-linear absorption. In addition, an interaction between the expanded plume containing B atoms and ambient gas takes place in the scattering region that is distanced from the hBN target. Its contribution to the surface nucleation is relatively low.

The source of nitrogen atoms may be dissociation of the adsorbed N₂ molecules by laser photons due to a lower-order non-linearity. When a molecule is adsorbed on the surface its forbidden electronic transitions may become allowed. The relevant two-photon transition of N₂ at λ_L≈248 nm responsible for the nitrogen atoms generation is $a^1\Pi_g \leftarrow X^1\Sigma_g^+$ ($v'-v''=7-0$) (Lyman-Birge-Hopfield system). Moreover, high vibrational levels v'>6 of the $a^1\Pi_g$ state are known to predissociate into the ground-state $N(^4S)$ atoms what may explain (8).

The produced atoms then recombine in three-body collisions with the ambient gas with the rate constant k≈10$^{-32}$ cm$^6$/s [33] :

$$N + N + N_2 \xrightarrow{k} N_2 + N_2 \qquad (9)$$

As a result, their instantaneous concentration after the laser pulse reduces according to a hyperbolical law with time constant $\tau^* = (kn_0[N_2])^{-1}$. The characteristic decay time of $\tau^* = 1$ μs from Eq. (7) corresponds to the atomic nitrogen population $n_0 = (k\tau^*[N_2])^{-1} \approx 10^{16}$ cm$^{-3}$. This concentration at the sample interface can be readily achieved in our experimental conditions.

Excited methastable BN molecules may also affect crystallization of nanorods. The potential curves of BN were theoretically calculated by Bauschlicher and Partridge [34]. The only excited electronic state accessible by one-photon transition from the $X^3\Pi$ ground state at hν=5.0 eV is $(3)^3\Pi$. This is a relatively long-lived state with lifetime of ~0.5 μs. Another longer-lived excited state is $(2)^1\Sigma^+$, which lifetime is $\tau^* \approx 2.6$ μs. It is positioned at 3.2 eV and cannot be excited directly. However, it can be populated from the $(3)^3\Pi$ state in course of the collisional relaxation. Because the collision-induced transitions to the low-lying Δ and triplet electronic states are forbidden in dipole approximation, the $(2)^1\Sigma^+$ quenching may be slow. Indeed, the spectral analysis of the expanding ablation plume shows two peaks at 393.6 nm and 397.1 nm (~3.1 eV), which may belong to the $(2)^1\Sigma^+ \to (1)^1\Delta, a^1\Sigma^+$ transitions [34]. Two lower-lying electronic states $(1)^3\Sigma^+$ and $(1)^3\Sigma^-$ with lifetimes correspondingly ~40 μs and ~10 μs may participate too. Their populations may be particularly high.

We remark that BN dissociation at the target cannot be a major source of N atoms. The last follows from Raman spectra of Fig. 2d, where no new band appears in presence of neon gas instead of nitrogen. For the same reason the metastable BN molecules cannot be the major source for the nanorods buildup.

Recently, Lee et al. [35] have reported successful synthesis of the BN nanotubes in gram quantities by $CO_2$-laser ablation in 1-bar nitrogen atmosphere. This method results in high target temperature of 3400 °C above that of the hBN decomposition ($T_0$). It was suggested that boron atoms agglomerate into hot clusters and nanoparticles and then react with gaseous $N_2$ forming BN nanotubes. This mechanism is however not appropriate to our experimental conditions because: (i) the growth takes place in the ablation region and (ii) the laser-matter interaction time is very short. The boron atoms can only aggregate between laser pulses as they are easily ablated under irradiation prior to hBN because of much higher absorption. On the other hand, during the time interval between successive laser pulses the reaction of boron with molecular nitrogen is expected to be prohibited because of a low temperature. Indeed, the heat diffusion length [9] at the characteristic backscattering time $t_{TOF} \approx 1$ μs (when scattered products impinge the target) is $l_T \approx 2\sqrt{Dt_{TOF}} = 40$ μm (here the hBN heat diffusivity $D = \chi/(\rho \cdot c_p) = 3.2$ cm$^2$/s). Because the modified region depth at single-short exposition is small $\delta \approx 1$ μm ($\delta \ll l_T$), the target temperature becomes low at this time: $T_{t=t_{TOF}} \ll T_0$. As a consequence of that, the gas interacts with a relatively cold surface. The nitrogen molecules do not react with boron atoms or solids in these conditions.

According to Golberg et al. [36] the BN plume interaction with nitrogen gas strongly contributes to the nanostructures formation. Pereira et al. [37] have recently discussed a backscattering of the ablated products in form of clusters, which may also be considered as elementary building units for the observed nanorods. The spatial distribution of these clusters is generally not uniform and mostly concentrated at the perimeter of the ablated area, which is explained by vortices formation favoring transversal mass transport. Indeed, an indication of this mechanism can be observed in Fig. 1(a,c). However, this model cannot explain the most intense structuring of the central part of the irradiated area in Fig. 1b evidenced in the present study.

We believe that in our experimental conditions the new solid phase is created by boron reaction with backscattered nitrogen atoms. High purity of this new phase disables its ablation allowing the nanorods to grow by consuming impurity-rich hBN. However, at laser fluences $E_L \geq 3E_a$ both old and new hBN phases are ablated resulting in a weak surface structuring. The fact that there exists optimum laser fluence $E_L > E_a$ for the structure appearance supports our conclusion about higher purity of the recrystallized product.

## 5. Conclusion

A multiple nucleation of a new crystalline product - BN nanorods - has been observed at UV-laser (248 nm) irradiation in the ablation regime ($E_L \geq E_{thr}=80$ mJ/cm$^2$) of hexagonal boron nitride (hBN) at moderately high ambient gas pressures above 500 bar. The effect is produced (1) in ablation regime, (2) under nitrogen gas pressure, (3) at low oxygen pressure below 0.2 mbar, and (4) it is specific to the laser pulse duration: the nanorods appear when it is shorter or equal to 5 ps. Moreover, the surface structuring is stronger when 5 ps pulse duration is employed. In contrast, at 150-ps laser irradiation or at fluences below the ablation threshold only amorphous boron suboxides $B_xO_y$ are observed.

Because of the sub-bandgap irradiation, the process is supposed to trigger on structural defects, which number density strongly decreases after recrystallization. Oxygen gas and generated nitrogen atoms efficiently heal nitrogen vacancies, which serve to be centers of the sample depletion. High-pressure ambient gas is responsible for the backscattering of the ablation products. We argue that electronically excited states contribute to the observed effect.

**Acknowledgments.** The work performed at ULF-FORTH has been supported through the access activities of the EC FP6 project "Laserlab-Europe" RII3-CT-2003-506350. The fluorescence spectroscopy analysis of hBN samples has been carried out at Hasylab (Desy) with a support by the RII3-CT-2004-506008 (IA-SFS) contract of the European Community. The authors are indebted to A. Egglezis with invaluable technical support and P. Jaffrennou for discussion of TEM images.

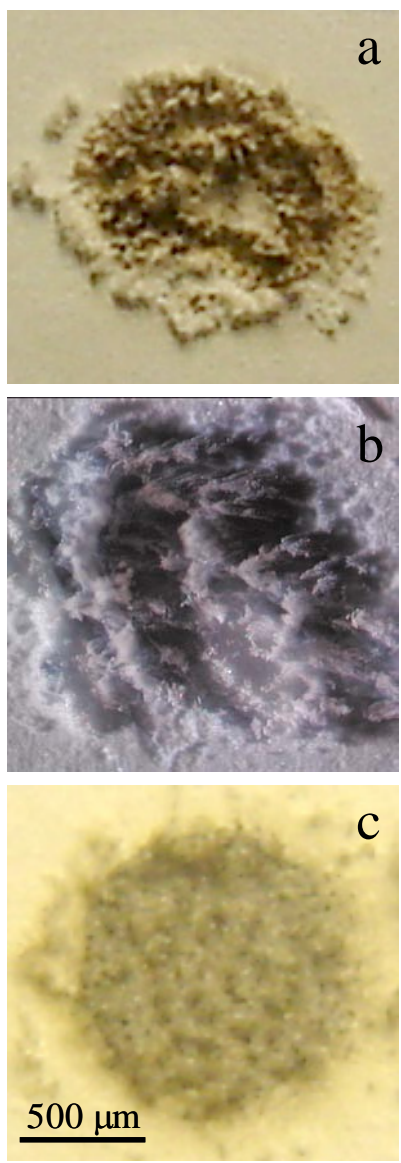

**Figure 1.** View of hBN sample after laser irradiation with pulse duration of 150 ps (a) 5 ps (b) and 0.45 ps (c). Ambient gas is $N_2$ at P=500 bar with traces of $O_2$: P~1-0.1 mbar. $E_L$=150 mJ/cm$^2$, dose~10$^3$ J/cm$^2$, $\lambda_L$=266 nm (a) and 248 nm (b-c).

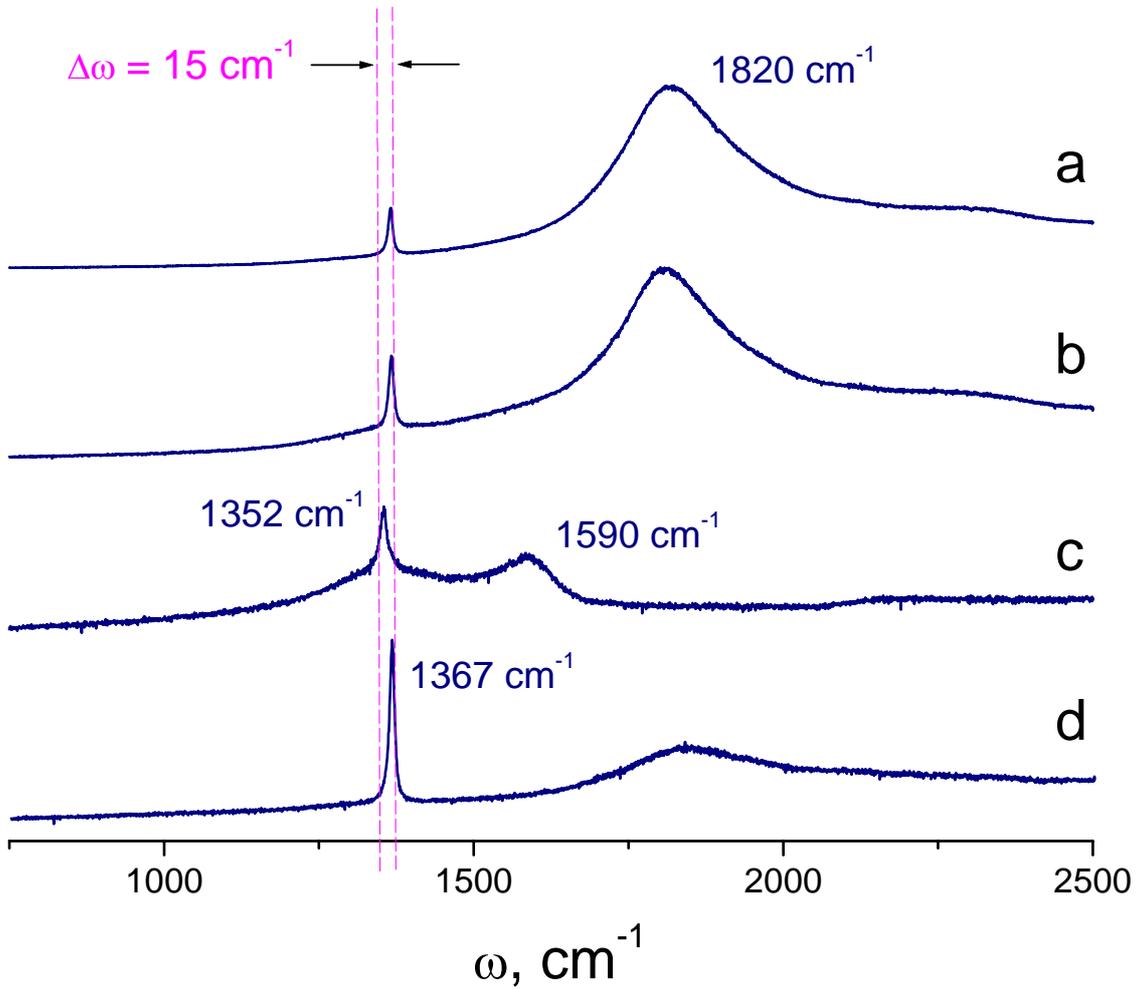

**Figure 2.** Raman spectra (488 nm, 40 mW) of the laser-irradiated hBN samples. The irradiation conditions are: 150 ps / 266 nm, (a) and 0.45 ps / 248 nm (b-d); the laser fluence is above (a,c-d) and below (b) the ablation threshold. The ambient gas is $N_2$ at 500 bar (a-c) with oxygen content of 20 mbar (a) and <0.2 mbar (b-c). Ne gas at 500 bar pressure is used instead of $N_2$ in (d).

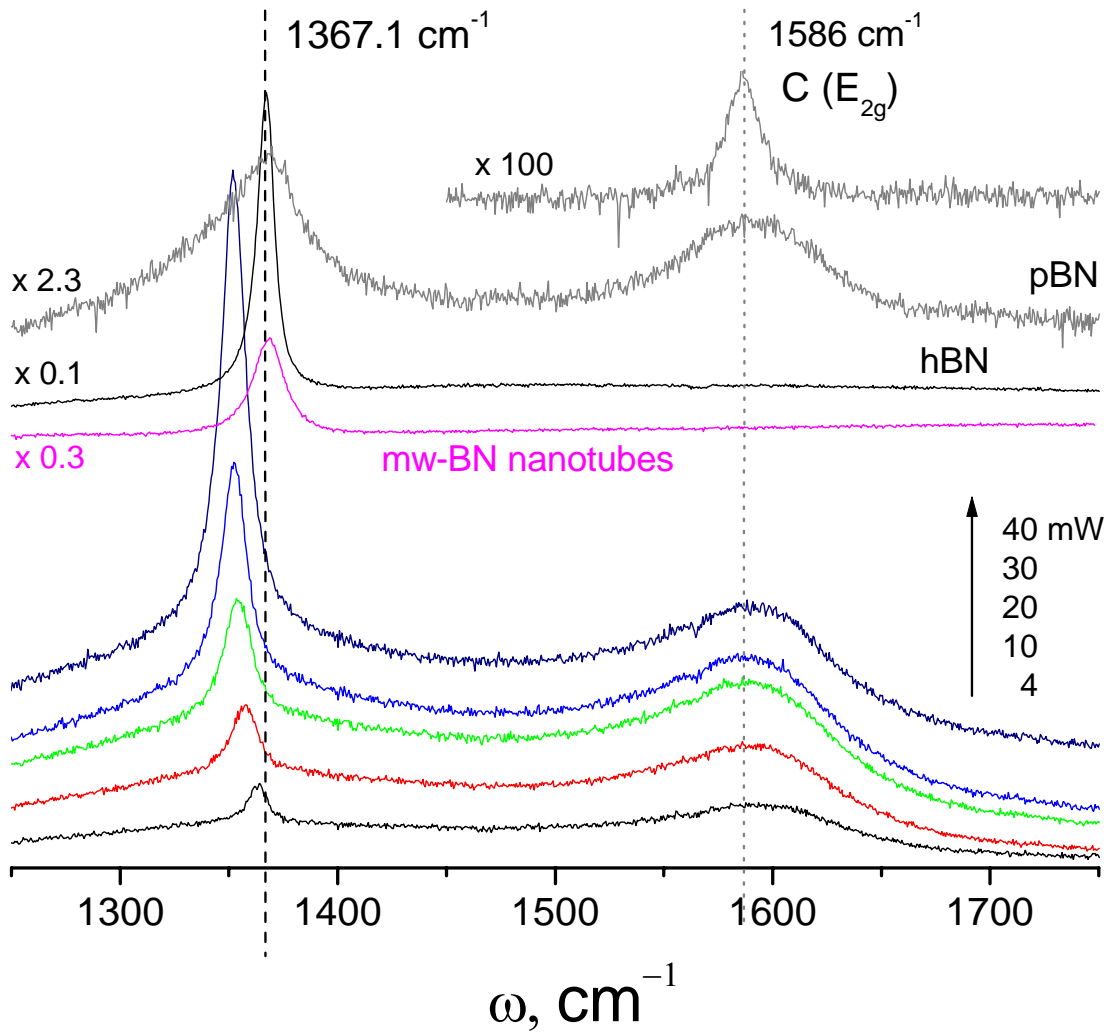

**Figure 3.** Raman spectra (488 nm) of the irradiated hBN samples (0.45 ps, 248 nm, $E_L$=150 mJ/cm$^2$, ambient gas $N_2$ at 500 bar, oxygen pressure <0.2 mbar), taken at cw-Ar$^+$ laser power varied from 4 mW to 40 mW. Four reference spectra of bulk hBN, pyrolytic BN (pBN), multiwall BN nanotubes, and graphite (C) taken at 40 mW cw-laser power are presented in the upper part of the figure.

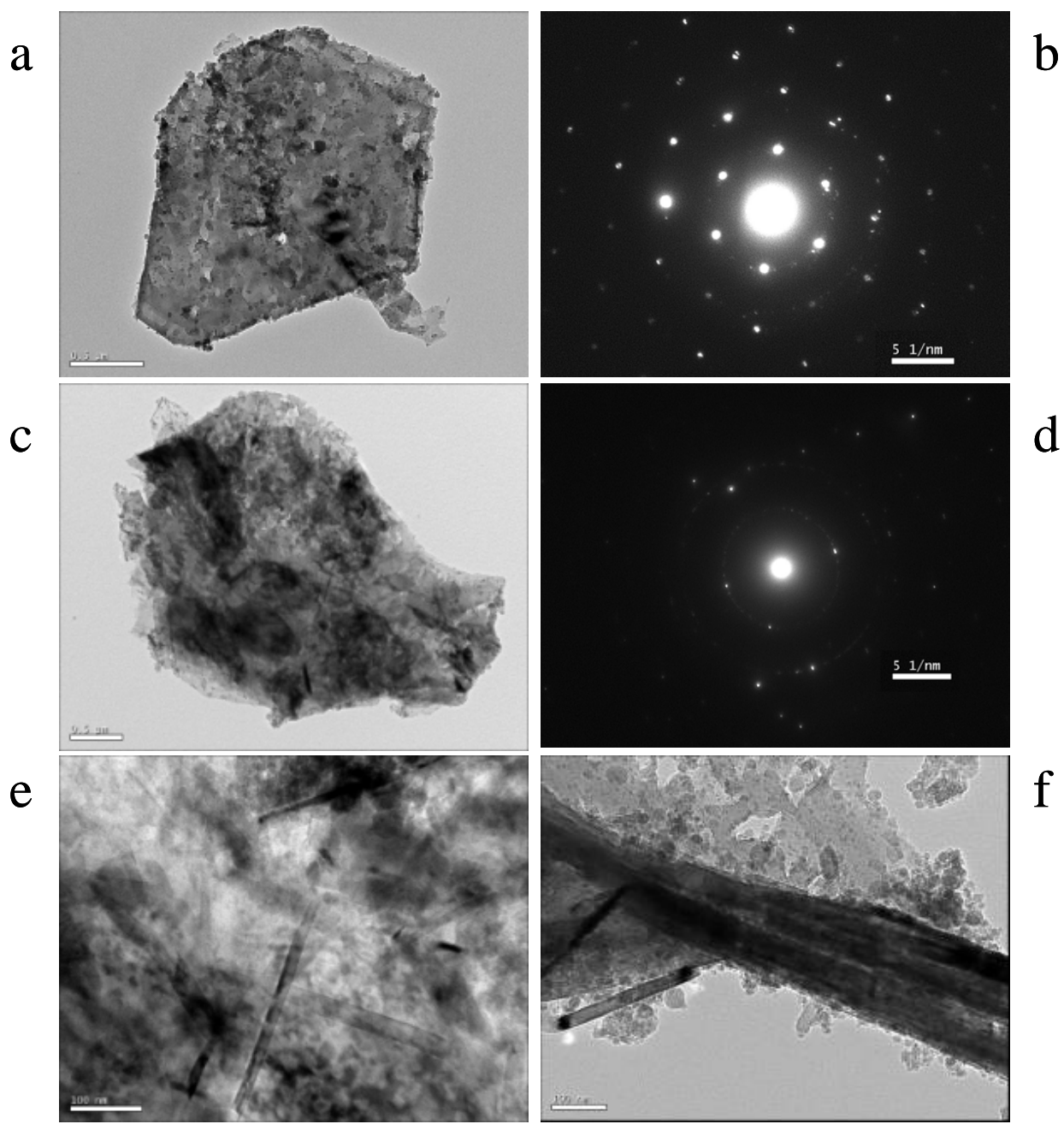

**Figure 4.** TEM images (a, c, e-f) and electron diffraction patterns (b, d) of fresh (a-b) and irradiated (c-f) hBN samples. The irradiation condition are: 5 ps / 248 nm, $E_L$=150 mJ/cm$^2$, dose~$10^3$ J/cm$^2$; ambient gas is $N_2$ at P=500 bar with traces of $O_2$ (<0.2 mbar).

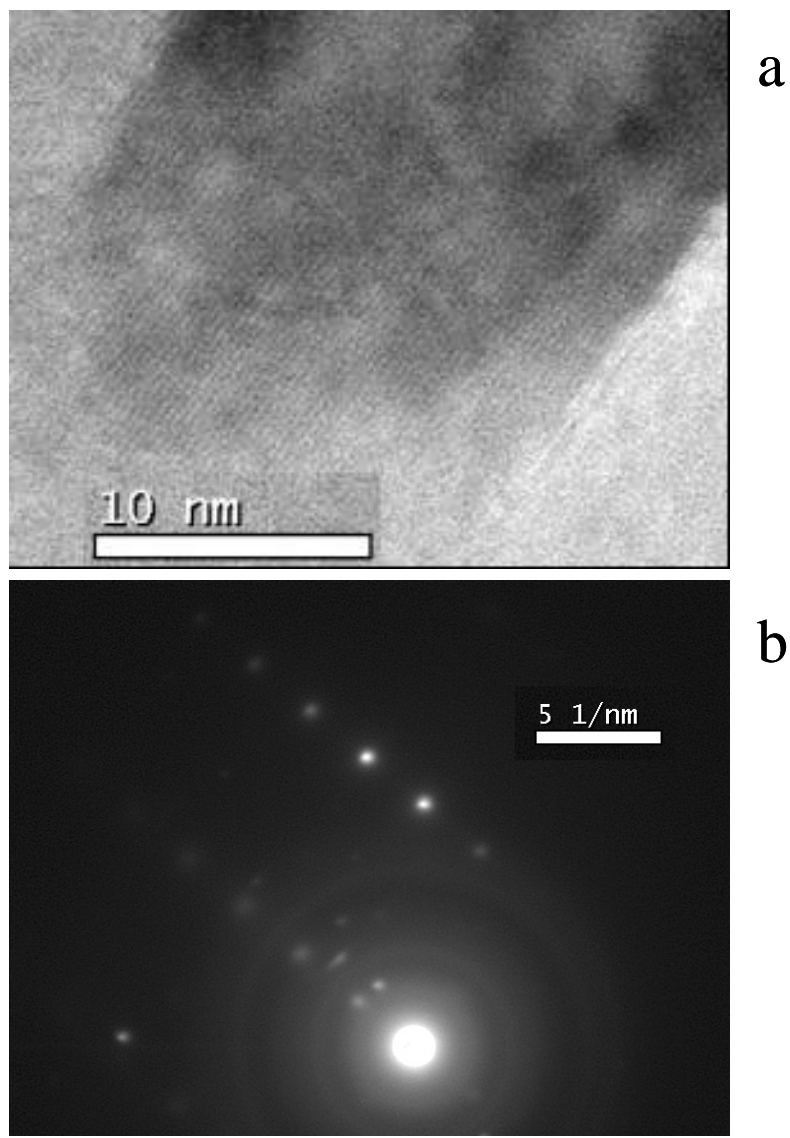

**Figure 5.**     TEM image (a) and electron diffraction pattern (b) of the hBN nanorod appeared in the irradiated area (irradiation condition as presented in Fig. 4).

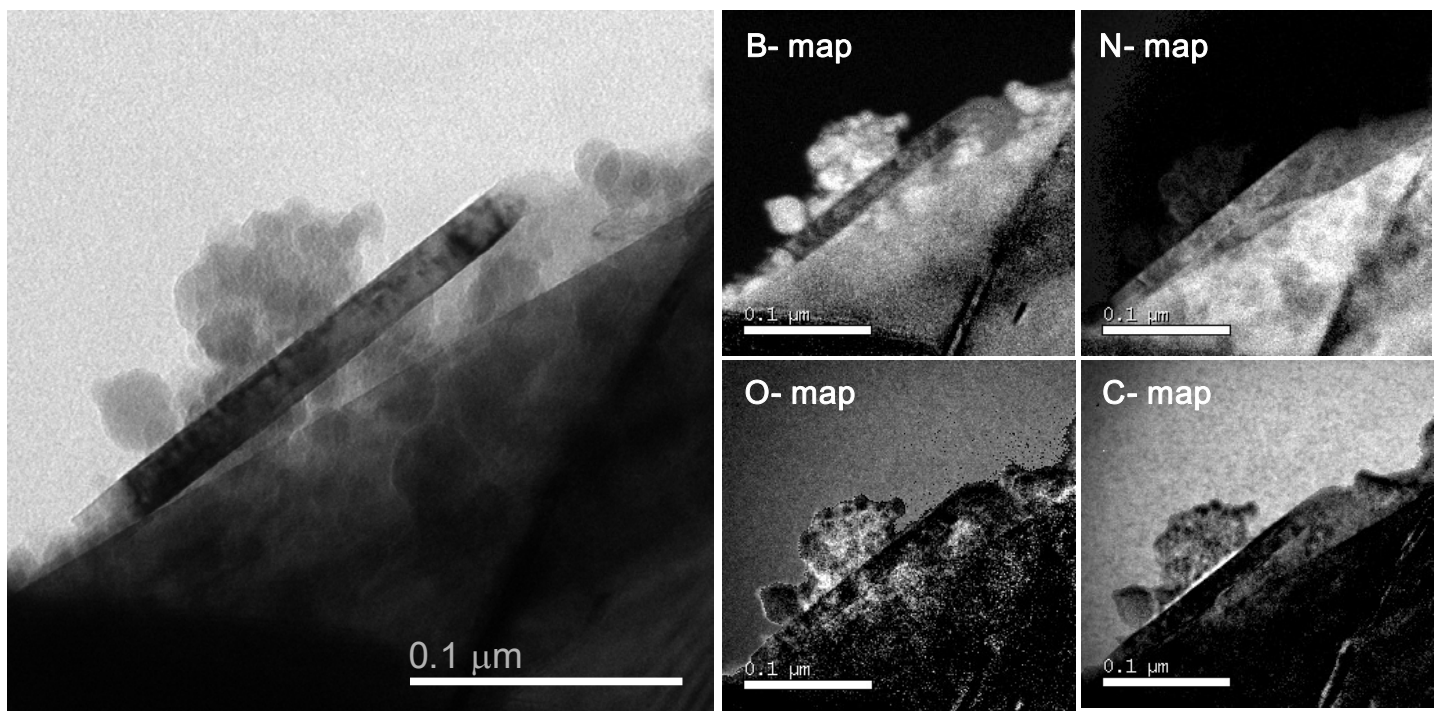

**Figure 6.**     EELS analysis of hBN nanorod:  (irradiation condition as presented in Fig. 4).

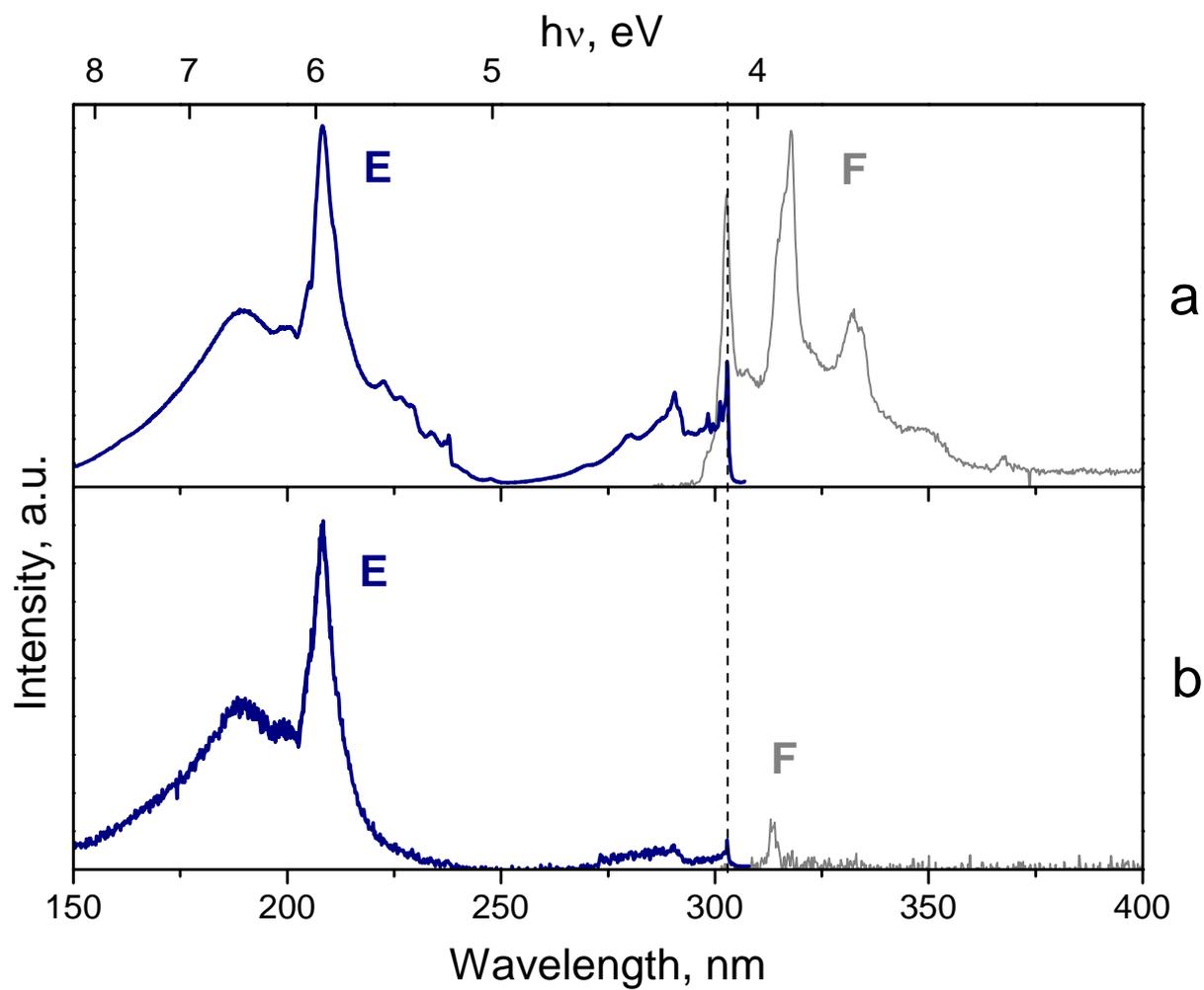

**Figure 7.** Fluorescence excitation spectra (E: $\lambda_{fluo}$=317 nm, not corrected for the primary monochromator efficiency) and fluorescence spectra (F: $\lambda_{exc}$=280 nm) of the original (a) and transformed (b) hBN samples (T=9 K).